\documentclass[fleqn,usenatbib]{mnras}
\usepackage{newtxtext,newtxmath}
\usepackage{mathptmx}
\usepackage[T1]{fontenc}
\DeclareRobustCommand{\VAN}[3]{#2}
\let\VANthebibliography\thebibliography
\def\thebibliography{\DeclareRobustCommand{\VAN}[3]{##3}\VANthebibliography}
\usepackage{graphicx}   
\usepackage{amsmath}    
\title[Fast-variation pulsations]{Detecting fast-variation pulsations in solar hard X-ray and radio emissions}

\author[D. Li]{Dong~Li,$^{1,2}$\thanks{E-mail: lidong@pmo.ac.cn}
\\
$^{1}$Purple Mountain Observatory, Chinese Academy of Sciences, Nanjing 210023, China \\
$^{2}$State Key Laboratory of Solar Activity and Space Weather, National Space Science Center, Chinese Academy of Sciences, Beijing 100190, China}
\date{Accepted 2025 June 08. Received 2025 June 03; in original form 2025 May 01}
\pubyear{2025}
\begin{document}
\label{firstpage}
\pagerange{\pageref{firstpage}--\pageref{lastpage}}
\maketitle
\begin{abstract}
Quasi-periodic pulsations (QPPs) at sub-second periods are
frequently detected in the time series of X-rays during stellar
flares. However, such rapid pulsations are rarely reported in the
hard X-ray (HXR) emission of the small solar flare. We explored the
QPP patterns with fast-time variations in HXR and radio emissions
produced in a small solar flare on 2025 January 19. By applying the
Fast Fourier Transform, the fast-variation pulsations at a
quasi-period of about 1~s are identified in the HXR channel of
20$-$80~keV, which were simultaneously measured by the Hard X-ray
Imager and the Konus-Wind. The rapid pulsations with a same
quasi-period were also detected in the radio emission at a lower
frequency range of about 40$-$100~MHz. The restructured HXR images
show that the QPP patterns mainly locate in footpoint areas that
connect by hot plasma loops, and they appear in the flare impulsive
phase. Our observations suggest that the fast-variation pulsations
could be associated with nonthermal electrons that are periodically
accelerated by the intermittent magnetic reconnection, and the 1-s
period may be modulated by the coalescence instability between
current-carrying loops and magnetic islands.
\end{abstract}

\begin{keywords}
Sun: flares -- Sun: oscillations -- Sun: X-rays -- Sun: radio
radiation
\end{keywords}
\section{Introduction}
The phenomenon of quasi-periodic pulsations (QPPs) is an
observed feature that is a highly variable modulation to the flare
emission, and it is always dominated by a series of regular,
repeated and successive pulsations in the time-intensity curve of
solar/stellar flares \citep[e.g.,][a reference therein]{Zimovets21}.
The QPPs behavior is a subject of much interest, because it is
important for us to understand the impulsive and periodic
energy-releasing processes in the solar atmosphere, i.e., the
intermittent magnetic reconnection, or the repetitive particle
acceleration \citep[e.g.,][]{Tan08,Yuan19,Li24a}. It was first
noted by \cite{Parks69} in wavebands of hard X-ray (HXR) and
microwave during a solar flare. And now, such quasi-periodic
behavior has been reported throughout all the electromagnetic
spectrum in wavebands of radio/microwave, white light, ultraviolet
(UV), extreme ultraviolet (EUV), soft and hard X-rays (SXR/HXR)
\citep[e.g.,][]{Tan16,Nakariakov18,Kolotkov21,Shen22,Motyk23,Li24b,Zhou24,Song25}.
Besides, the QPPs feature has been detected in wavelengths of
Ly$\alpha$ \citep{Lid22}, H$\alpha$
\citep{Kashapova20}, and $\gamma$-ray line at 2.223~MeV
\citep{Li22}. Although the QPPs behavior has been observed in a
large number of solar/stellar flares, the generation mechanism of
flare QPPs is still an open issue \citep{Kupriyanova20,Zimovets21}.
The flare QPPs are often associated with magnetohydrodynamic (MHD)
waves \citep{Nakariakov20}, particularly that the flare QPPs are
highly related to the magnetic loop and current sheet
\citep[cf.][]{Zimovets21}. The flare QPPs can also be triggered by
the intermittent magnetic reconnection. The idea is that nonthermal
particles could be periodically accelerated by the quasi-periodic
reconnection \citep{Zhang24,Lid25}. This model is well to produce
the periodic pulsations that are observed in HXR and microwave
channels during the flare impulsive phase
\citep[e.g.,][]{Clarke21,Li21,Kumar25}. However, there is not a
mechanism to explain all the observed QPPs, which may be attributed
to the observational fact that the current data can not provide
sufficient information to distinguish various models \citep{Kupriyanova20}.

The term `QPPs' generally refers to the presence of at least three
successive and periodic pulsations in flare intensity curves
\citep[e.g.,][]{McLaughlin18}, although this is not a strict
definition. On the other hand, the term `quasi-period', which is a
varying instantaneous period, is usually used for flare QPPs
\citep{Nakariakov19}. Since the first report on flare QPPs, their
quasi-periods have been detected in a wide range of timescales,
i.e., subseconds, several to dozens of seconds, and even hundreds of
seconds
\citep[e.g.,][]{Carley19,Knuth20,Altyntsev22,Collier23,Zhao23,French24}.
It appears that the observed quasi-periods strongly depend upon
wavebands. For instance, the quasi-periods at the order of
subseconds are mostly reported in wavebands of radio or microwave
\citep{Tan10,Karlicky24}, mainly because that solar telescopes often
have high temporal cadences in radio bands, i.e., with temporal
cadences of a few milliseconds. On the contrary, the flare QPPs that
are observed in SXR and HXR channels usually show longer
quasi-periods, i.e., $\geq$4~s \citep{Hayes20,Collier23}. Actually,
the flare QPPs at shorter quasi-periods (i.e., $\leq$1~s) have been
reported in the high-energy emission of HXRs and $\gamma$-rays, and
they are found to be produced in large (i.e., M- or X-class) solar
flares \citep{Knuth20,Li25}. Moreover, the subsecond QPPs were also
observed in X-rays during the stellar flare \citep{Misra20},
suggesting that such rapid pulsations tend to appear in large
flares. The newly launched Hard X-ray Imager \citep[HXI;][]{Su19} on
board the Advanced Space-based Solar Observatory (ASO-S) has a high
temporal cadence in the burst mode, providing us with an opportunity
to investigate the fast-variation pulsations in HXR emissions and
localise their generated sources. In this article, we investigated a
short-period QPP in HXR and radio emissions during a small flare.

\section{Observations}
We analysed a C-class solar flare that was located in the active
region of NOAA~13959. It started at $\sim$16:58~UT, peaked at
$\sim$17:08~UT, and stopped at about 17:12~UT. The solar flare was
measured by the Konus-Wind \citep[KW;][]{Lysenko22}, ASO-S/HXI, the
Spectrometer/Telescope for Imaging X-rays \citep[STIX;][]{Krucker20}
for the Solar Orbiter, the Geostationary Operational Environmental
Satellite (GOES), the Atmospheric Imaging Assembly
\citep[AIA;][]{Lemen12} on board the Solar Dynamics Observatory
(SDO), the Chinese H$\alpha$ Solar Explorer \citep[CHASE;][]{Li19},
the STEREO/WAVES (SWAVES), the Expanded Owens Valley Solar Array
(EOVSA), and the e-CALLISTO radio spectrograph at Greenland.

KW can provide the count rate light curves in three wide energy
channels with accumulation time of 2.944~s in the waiting mode, and
it switches to the varying temporal cadences (e.g., 2$-$256~ms) in
the triggered mode. ASO-S/HXI takes the imaging spectroscopy of
solar flares in the high-energy range of about 10$-$300~keV. Its
temporal cadence is normally 4~s in the regular mode and changes to
be 0.25~s in the burst mode. STIX measures the solar flare in the
X-ray range of 4$-$150~keV, and the scientific data at level~1
(i.e., L1) for this event has a temporal cadence of about 3~s. GOES
can provide full-disk light curves in SXR channels of 1$-$8~{\AA}
and 0.5$-$4~{\AA} at an uniform cadence of 1~s.

SDO/AIA takes snapshots over the entire Sun in multiple UV/EUV
wavelengths. In this study, we used the synoptic images, which has a
temporal cadence of 2~minutes, and a pixel scale of 2.4\arcsec.
CHASE provides the spectroscopic data of the entire Sun in the
waveband of H$\alpha$, and it has a pixel scale of 1.04\arcsec, and
a temporal cadence of about 71~s. SWAVES measures the solar dynamic
spectrum in the frequency range of about 0.00261$-$16.025~MHz, and
it has a temporal cadence of 1~minute. EOVSA is a solar
radioheliograph, providing the radio dynamic spectrum in frequencies
of $\sim$1$-$18~GHz, the temporal cadence can be as high as 1~s. It
can also provide microwave maps with a spatial scale of
2.5\arcsec~pixel$^{-1}$. The e-CALLISTO radio spectrograph at
Greenland provides the radio dynamic spectrum in the
frequency range of $\sim$10$-$106~MHz, which has an uniform cadence
of 0.25~s.

\section{Methods and Results}
\begin{figure}
\centering
\includegraphics[width=\columnwidth]{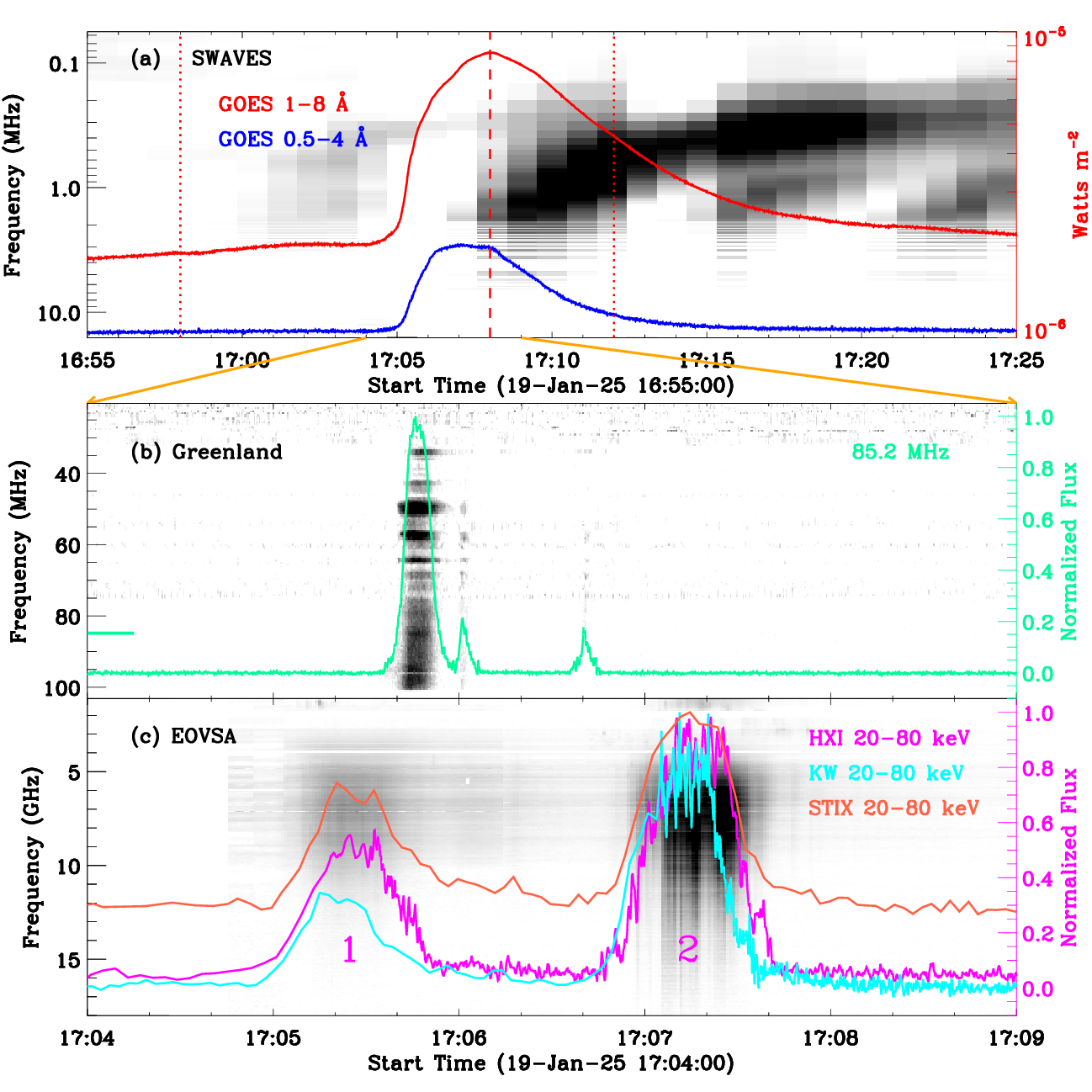}
\caption{Overview of the C8.2 flare on 2025 January 19. (a) SXR
fluxes recorded by GOES at 1$-$8~{\AA} (red) and 0.5$-$4~{\AA}
(blue). The vertical lines marks the start, peak and stop time of
the C8.2 flare. (b) Radio time series at the frequency of 85.2~MHz
(spring green). (c) Normalized HXR light curves captured by
HXI~20$-$80~keV (magenta), KW~20$-$80~keV (cyan) and
STIX~20$-$80~keV (tomato). The backgrounds are radio dynamic spectra
measured by SWAVES, e-Callisto, and EOVSA respectively.
\label{over}}
\end{figure}

Figure~\ref{over} shows an overview of the C8.2 flare in multiple
wavebands. Panel~(a) presents the GOES fluxes in SXR energy bands of
1$-$8~{\AA} and 0.5$-$4~{\AA} from 16:55~UT to 17:25~UT. The
background is the radio dynamic spectrum measured by
SWAVES, suggesting that a group of type III radio bursts is
accompanied by the C-class solar flare. The radio bursts in the
frequency range of about 20-100~MHz are also measured by the
e-CALLISTO spectrograph at Greenland, and they appeared in
the rise phase of the solar flare, as shown in panel~(b). The
overplotted curve is the radio flux at the frequency of 85.2~MHz.
Figure~\ref{over}~(c) presents three intensity curves in HXR energy
bands of HXI~20$-$80~keV, KW~20$-$80~keV, and STIX~20$-$80~keV
during 17:04$-$17:09~UT. In this study, the HXI intensity curve is
extracted from an open flux monitor and has a higher cadence of
0.25~s, the KW flux has been interpolated into an uniform cadence of
0.128~s, and the STIX flux is derived from the science pixelated
data. The three HXR intensity curves show two main peaks, as marked
by the Arabic numerals. The two main peaks can also be found in the
microwave emission measured by EOVSA. We can note that a number of
repetitive and successive wiggles superimposed on the main peak~2,
which might be regarded as the fast-variation pulsations, terming as
`short-period QPP'. It is clearly observed in the intensity curves
from ASO-S/HXI and KW, but not measured by STIX due to its
3-s cadence. We also note that the short-period QPP is not seen
during the main peak~1, which can be attributed to the normal mode
of HXI and KW.

Based on the Lomb-Scargle periodogram method \citep{Scargle82},
the Fast Fourier Transform (FFT) is utilized to the raw
intensity curves in wavebands of HXR and radio, and their Fourier
power spectra are derived. Next, a simple model ($S$) is applied for
fitting the Fourier power spectrum, that is, a power-law
distribution pluses a constant ($C$) term
\citep[e.g.,][]{Liang20,Anfinogentov21}, as shown in
Eq.~(\ref{eq1}).
\begin{equation}
\centering
 S (P) = A~P^{-\alpha} + C,
\label{eq1}
\end{equation}
\noindent Where $P$ is the period distribution of Fourier power
spectra, $A$ denotes to the amplitude of Fourier power spectra, and
$\alpha$ represents a power-law index.

\begin{figure}
\centering
\includegraphics[width=\columnwidth]{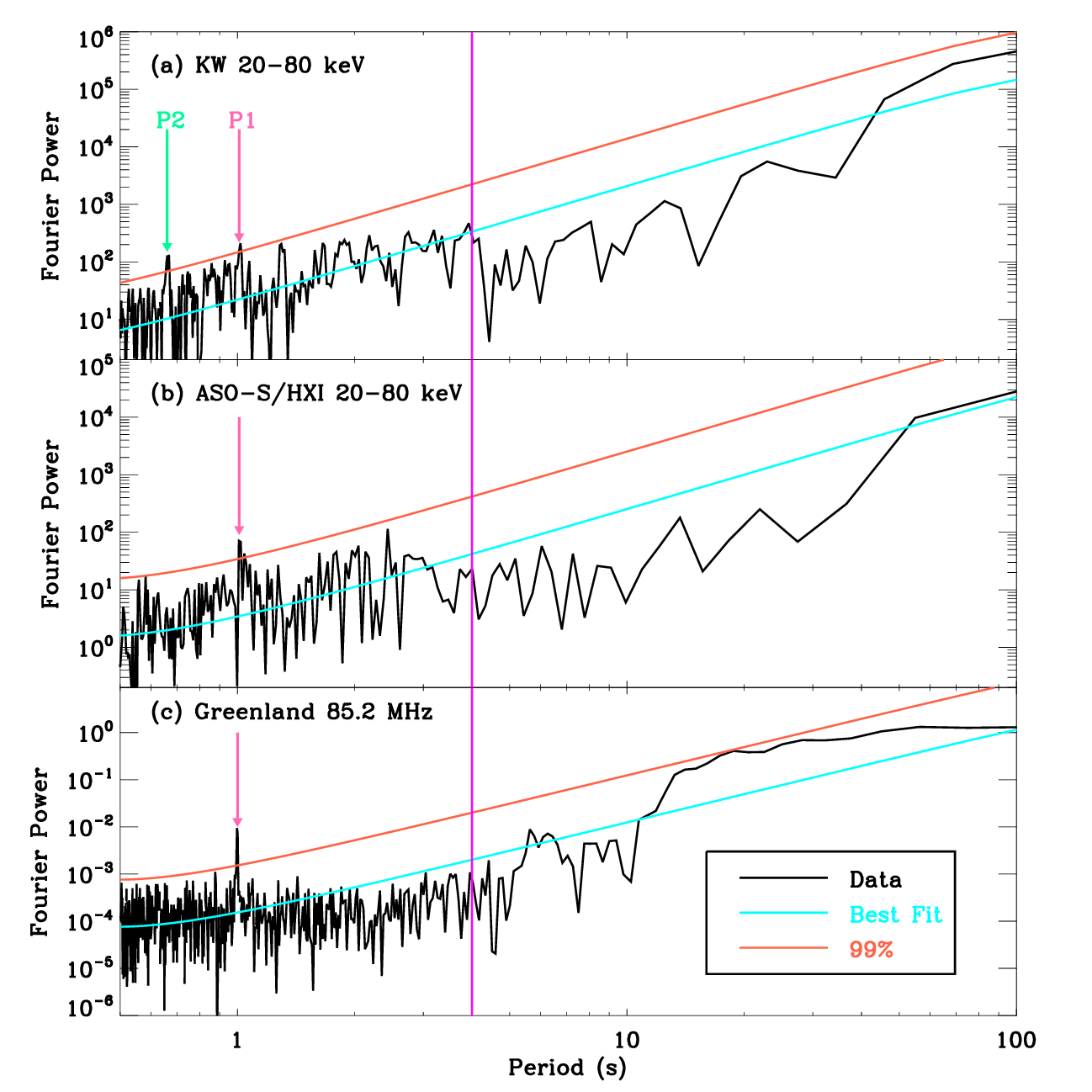}
\caption{Fourier power spectra during the C8.2 flare in wavelengths
of HXR (a \& b) and radio (c). The cyan line in each panel
represents the best fit for the observational data (black), the
tomato line is the confidence level at 99\%. The hot pink arrow
outlines the interested period above the 99\% confidence level.
\label{fft}}
\end{figure}

Figure~\ref{fft} shows the Fourier power spectra and their best-fit
results in high-energy HXR and low-frequency radio emissions. We
note that two peaks are above the 99\% confidence level in the HXR
energy range of 20$-$80~keV measured by KW, corresponding to double
quasi-periods at about 0.66~s and 1.0~s, as indicated by the hot
pink and green arrows. On the other hand, only one peak exceeds the
99\% confidence level in channels of HXI~20$-$80~keV and the
low-frequency radio emission, confirming the existence of
fast-variation pulsations with a quasi-period of about 1~s. The
shorter quasi-period at about 0.66~s is not detected by ASO-S/HXI
and e-CALLISTO, mainly limiting to their temporal cadence,
i.e., 0.25~s. The 1-s QPP can be detected in data of three different
instruments, such as two spacecrafts and one ground-based telescope,
which clearly exclude instrumental artefacts and demonstrate the
presence of fast-variation pulsations during the C-class flare.

\begin{figure}
\centering
\includegraphics[width=\columnwidth]{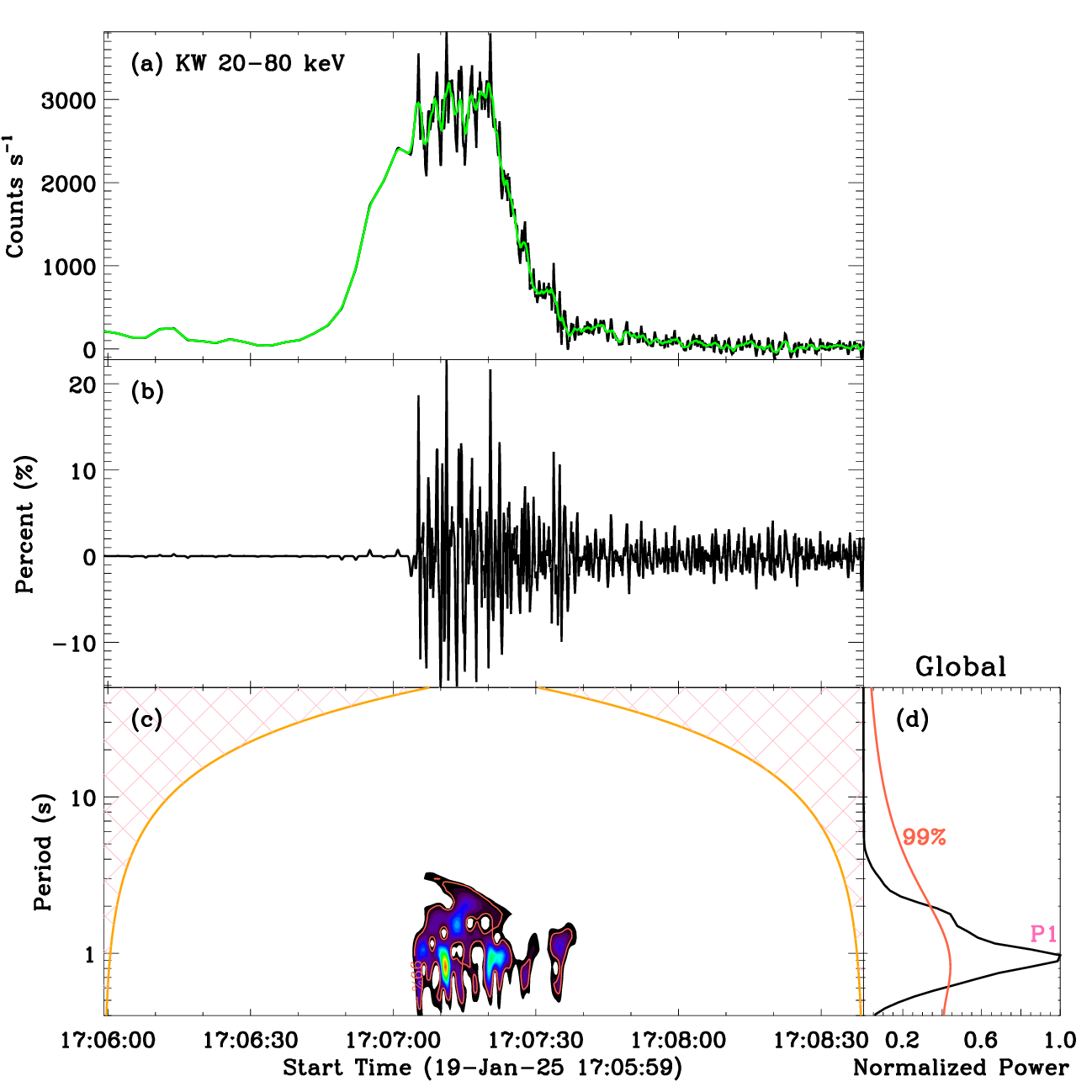}
\caption{Morlet wavelet analysis for the HXR flux of the C8.2 flare.
(a) Raw light curve at KW~20$-$80~keV. The overlaid green lines
represent its long-term trend. (b) Detrended time series normalized
to the maximum of the long-term trend. (c) Morlet wavelet power
spectrum for the detrended time series. (d) Global wavelet power
spectrum. The tomato contours or lines indicate a significance level
of 99\%. \label{wav}}
\end{figure}

In order to look closely at the fast-variation pulsations, the
wavelet analysis method with a mother function of `Morlet'
\citep{Torrence98} is used for the detrended time series in the
waveband of KW~20$-$80~keV, since it has a higher temporal cadence
of 0.128~s. The raw light curve is decomposed into a long-term trend
and a detrended time series by using the Fourier method. In our
case, a threshold timescale of 4~s is chosen to distinguish them, as
indicated by the vertical magenta line in Figure~\ref{fft}.

Figure~\ref{wav} presents the wavelet analysis results. Panel~(a)
draws the HXR flux and its long-term trend. Here, the long-term
trend can be regarded as the background emission, which matches the
raw light curve. Panel~(b) plots the detrended time series, which
has been normalized by its long-term trend. It clearly reveals the
QPP patterns at a shorter period when the KW instrument switches to
the triggered mode. The modulation depth, which can be determined by
the ratio between the detrended time series and its long-term trend,
is more than 10\% during these fast-variation pulsations. It can be
as high as 20\% in some rapid pulsations, while it is rather small
(i.e., $<$3\%) after the QPP patterns. Panels~(c) \& (d) show the
wavelet power spectrum and its Global wavelet power spectrum. They
are both characterized by a bulk of power spectrum inside the 99\%
confidence level, which are centered at about 1~s. Also, these two
power spectra show a wide period range, which may be attributed to
the lower resolution of the wavelet analysis method, and this
matches two quasi-periods obtained from the FFT method in
Figure~\ref{fft}~(a). The 1-s QPP patterns appear during the flare
impulsive phase, i.e., roughly from 17:07:05~UT to 17:07:35~UT. All
these observations demonstrate that the short-period QPP comes from
the solar flare rather than the solar background emission.

\begin{figure}
\centering
\includegraphics[width=\columnwidth]{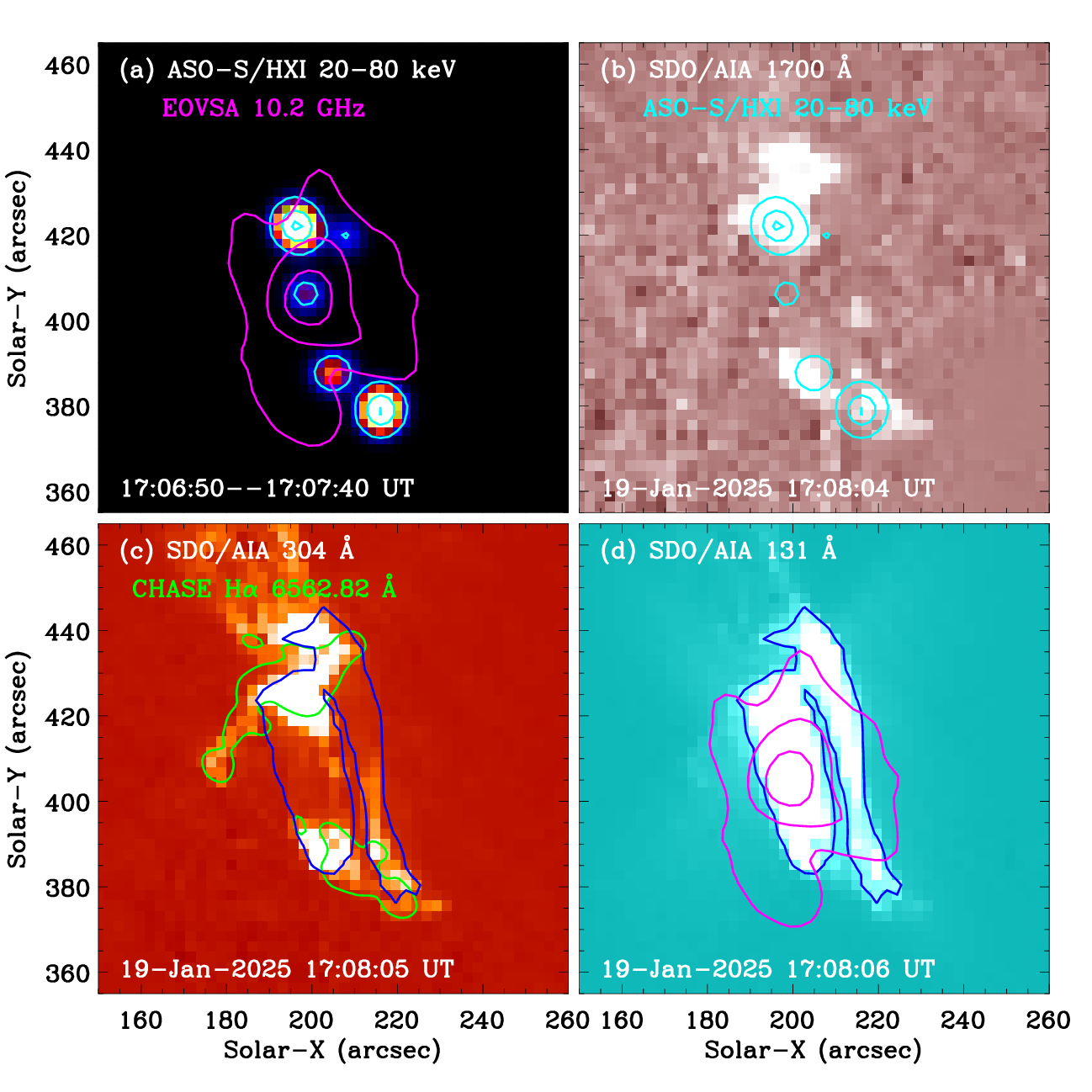}
\caption{Multi-wavelength images with a FOV of
$\sim$110\arcsec$\times$110\arcsec\ during the C8.2 flare. (a) HXR
map in the energy range of HXI~20$-$80~keV. The contours represent
the HXR (cyan) and microwave (magenta) emissions, respectively, and
their contour levels set at 10\%, 50\%, and 90\% in the HXR channel,
while those are 20\%, 50\%, and 80\% in the microwave emission. (b)
UV map at AIA~1700~{\AA}. (c \& d) EUV maps in wavelengths of
AIA~304 and 131~{\AA}. The color contours outline flare ribbons at
H$\alpha$~6562.82~{\AA} (green) and hot loops at AIA~131~{\AA}
(blue), respectively. \label{smap}}
\end{figure}

Figure~\ref{smap} shows the spatial structure of the C8.2 flare in
multiple wavelengths of HXR, microwave, H$\alpha$, and UV/EUV.
Panel~(a) draws a HXR map in the high-energy range of 20$-$80~keV,
the overplotted contours represent the HXR (cyan) and microwave
(magenta) radiation. In this case, the HXR map is reconstructed from
the ASO-S/HXI data with a algorithm of HXI\_CLEAN, and it exhibits
several HXR radiation sources, which match those UV bright kernels
in the photosphere seen in the wavelength of AIA~1700~{\AA}, as
shown in panel~(b). These HXR sources or UV kernels may be regarded
as the conjugate points that connected by hot plasma loops, and the
loop-top source can be seen in passbands of microwave or HXR, as
indicated by the magenta contour. Panel~(c) \& (d) shows the EUV
maps in wavebands of AIA~304~{\AA} and 131~{\AA}, which contain the
lower ($\sim$0.05~MK) and higher ($\sim$10~MK) temperature plasmas,
respectively. Some ribbon-like structures are observed in wavebands
of AIA~304~{\AA} and CHASE~H$\alpha$~6562.82~{\AA}, and they are
overlapping on the UV kernels or HXR sources, indicating that those
ribbon-like structures are flare ribbons. Two loop-like features can
be clearly seen in the AIA~131~{\AA} map, and they connect those
flare ribbons or kernels through the loop-top region, as outlined by
the blue, green, and magenta contours. These observations suggest
that the main HXR sources could be regarded as footpoints, and the
strongest microwave emission may locate at the loop-top source. The
HXR sources at footpoints may contain a series of magnetic islands
that are formed in opposite magnetic polarities. Assuming a
semi-circular profile of flare loops \citep[cf.][]{Li22}, the loop
length ($L$) may be estimated by the distance between the flare
ribbons or kernels, which is in the range of 40$-$65~Mm.

\section{Conclusion and Discussion}
We present an observational detection of 1-s QPP of HXR and radio
emissions produced in a small solar flare on 2025 January 19. The
fast-variation pulsations in channels of HXR and low-frequency radio
are detected by ASO-S/HXI, KW, and e-CALLISTO, mainly
benefitting from the high-cadence burst mode of the three
telescopes. Combining the Fourier Transform
\citep{Anfinogentov21} and wavelet analysis
\citep{Torrence98} method, the quasi-period centered at about 1~s is
simultaneously identified in passbands of KW~20$-$80~keV,
HXI~20$-$80~keV, and Greenland~85.2~MHz. The presence of
fast-variation pulsations in data of three different instruments on
two spacecrafts and one ground-based telescope clearly excludes the
instrumental artefact. The modulation depth of the 1-s QPP is in the
range of about 10\%$-$20\%, which is close to that of flare QPPs at
longer periods in HXRs and $\gamma$-rays \citep{Li22}.

Generally, the very-short period pulsations (i.e., $\ll$1~s) were
often reported in solar radio emissions, especially in the microwave
regime \citep[e.g.,][]{Tan10,Nakariakov18,Karlicky24}. Conversely,
the flare QPPs in high-energy emissions are usually found to show
the characteristic periods above 4~s
\citep[e,g.,][]{Parks69,Li22,Collier23,Inglis24}, which
may be attributed to the observational limitation, for instance, the
signal-to-noise ratio of solar telescopes in channels of HXRs or
$\gamma$-rays is a bit low. By statistically investigating the solar
flares measured by Fermi/GBM in the high-cadence burst mode,
\cite{Inglis24} demonstrate that the short-period QPPs at the
timescale of about 1$-$4~s are actually existing in the HXR channel,
but they are not widespread. Recently, the fast-variation pulsations
at quasi-periods that is not more than 1~s have been detected in the
high-energy range of HXR and $\gamma$-ray continuum during large
solar flares, i.e., a M-class \citep{Knuth20} and an X-class flares
\citep{Li25}. In our case, the fast-variation pulsations at a
quasi-period of about 1~s is simultaneously observed in passbands of
HXR and low-frequency radio emissions during a small flare.
Moreover, the spatial structure of the QPP patterns is reconstructed
using the ASO-S/HXI data, providing us an opportunity to explore
their generation mechanism.

\begin{figure}
\centering
\includegraphics[width=\columnwidth]{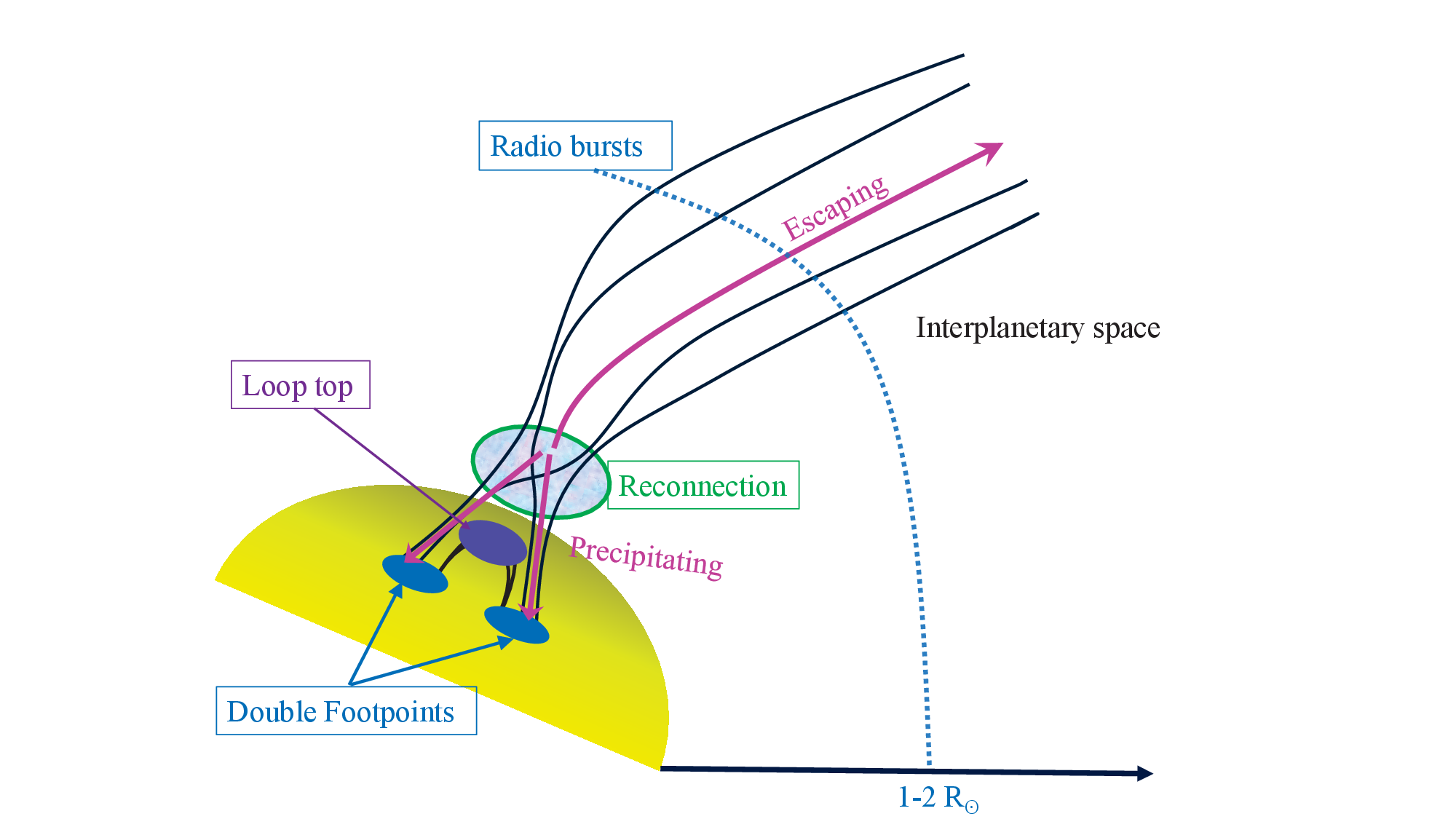}
\caption{Cartoon of the flare and the type III radio bursts, which
is used for illustrating the likely mechanism that was observed the
particle acceleration resulting in QPPs in radio/microwave and HXR
emissions. The QPP source regions such as the loop top and double
footpoints are related to the open and closed field lines that
allows for the escaping (or upward) and precipitating (or downward)
electrons. \label{cart}}
\end{figure}

The QPP patterns are frequently interpreted as MHD waves
\citep[e.g.,][]{Nakariakov20}. Here, the phase speed ($v_P$) may be
determined by the flare loop length ($L$) and the quasi-period
($P$), which is $v_P=2L/P \approx (0.8-1.3) \times
10^5$~km~s$^{-1}$. The phase speed is approximately equal to
(0.27-0.43)$c$ ($c \approx 3 \times 10^5$~km~s$^{-1}$ is the speed
of light), and it is much faster than the local sound and Alfv\'{e}n
speeds. Thus, the short-period QPP is difficult to be modulated by
the MHD wave. We also note that the 1-s QPP tends to appear in the
impulsive phase of the C8.2 flare, and it can be simultaneously seen
in HXR and low-frequency radio emissions, implying the existence of
nonthermal electrons. Therefore, the flare QPP might be triggered by
a periodic regime of magnetic reconnection, which may periodically
accelerate nonthermal electrons, as shown in Figure~\ref{cart}.
Briefly, the pulsed nonthermal electrons are accelerated by the
magnetic reconnection and show bi-directional propagations, as
indicated by the magenta arrows. The downward electron precipitates
toward the lower atmosphere along the closed magnetic field line,
leading to the HXR or microwave pulsation at double footpoints or
loop-top regions. On the other hand, the upward electron escapes
along the open magnetic field line and produces the low-frequency
radio pulsation in the interplanetary space, i.e., type III radio
bursts. Such reconnection process occurs repeatedly and regularly,
generating the fast-variation pulsations at a quasi-period in HXR
and low-frequency radio emissions \citep{Clarke21,Li21}. The short
quasi-period at about 1~s might be explained by the current-loop
coalescence model presented by \cite{Tajima87}, who have
demonstrated that the timescale of transit pulsations in the range
of 1$-$10~s could be modulated by the coalescence instability of
current-carrying loops and magnetic islands. In this model, the
magnetic islands are located in current sheets that are generated by
the interaction of various magnetic loops with opposite magnetic
polarities. The coalescence instabilities between magnetic loops and
current sheets with a quasi-period would cause periodic energy
releases in the solar flare \citep[e.g.,][]{Zhao23}, that is, a
periodic regime of magnetic reconnection occurs during the flare
impulsive phase. The detection of HXR sources indicates some
magnetic islands, and the observation of hot plasma loops rooted in
different footpoints suggests various magnetic loops with opposite
magnetic polarities, although the current sheets are not seen in our
case due to the observational limitation, i.e., the lower spatial
resolution. It has been demonstrated that the
current-carrying loop itself is easily exciting a tearing-mode
oscillation, modulating the magnetic reconnection process and
generating a QPP pattern at quasi-periods ranging from sub-seconds
to a few seconds \citep[cf.][]{Tan10b,Tan12}. Therefore, we may
conclude that the fast-variation pulsations at a quasi-period of 1~s
is probable associated with the intermittent magnetic reconnection
due to the coalescence of current-carrying loops.

\section*{Acknowledgements}
This work is funded by the National Key R\&D Program of China
2022YFF0503002 (2022YFF0503000) and 2021YFA1600502 (2021YFA1600500),
the Strategic Priority Research Program of the Chinese Academy of
Sciences, Grant No. XDB0560000. D.~Li is also supported by the
Specialized Research Fund for State Key Laboratory of Solar Activity
and Space Weather. We thank the teams of ASO-S/HXI, KW, STIX, GOES,
SDO, CHASE, and e-CALLISTO for their open data use policy.

\section*{Data Availability}
Publicly available data was analyzed and they can be found here:
~~http://aso-s.pmo.ac.cn/sodc/dataArchive.jsp,
~~https://www.ioffe.ru/LEA/kwsun/, ~~/http://jsoc.stanford.edu/,
~~https://datacenter.stix.i4ds.net/stix,
~~https://soleil.i4ds.ch/solarradio/callistoQuicklooks/.


\bsp    
\label{lastpage}
\end{document}